\documentclass[twocolumn,nofootinbib,
superscriptaddress,
amsmath,
amssymb,
aps,
prl,
]{revtex4-2}

\usepackage{graphicx}
\usepackage{xcolor}
\usepackage{dcolumn}
\usepackage{bm}
\usepackage{mathtools}
\usepackage{physics2}
    \usephysicsmodule{ab}
    \usephysicsmodule{ab.braket}
\usepackage{bibunits}
\renewcommand{\figurename}{Fig.}

\DeclarePairedDelimiter{\norm}{\lvert\!\lvert}{\rvert\!\rvert}

\DeclarePairedDelimiter{\lbra}{\langle\!\langle}{\rvert}
\DeclarePairedDelimiter{\lket}{\lvert}{\rangle\!\rangle}
\DeclarePairedDelimiterX{\rlbraket}[2]{\langle}{\rangle\!\rangle}{#1\,\delimsize\vert\,#2}
\DeclarePairedDelimiterX{\lrbraket}[2]{\langle\!\langle}{\rangle}{#1\,\delimsize\vert\,#2}
\DeclarePairedDelimiterX{\llbraket}[2]{\langle\!\langle}{\rangle\!\rangle}{#1\,\delimsize\vert\,#2}

\newcommand{\mqty}[1]{\begin{matrix}#1\end{matrix}}

\DeclarePairedDelimiter{\floor}{\lfloor}{\rfloor}

\definecolor{hrefcolor}{HTML}{039393}
\usepackage[
 colorlinks=true,
 allcolors=hrefcolor,
]{hyperref}

\begin{document}

\begin{bibunit}[apsrev4-2]

\preprint{APS/123-QED}

\title{Quasiperiodicity-induced non-Hermitian skin effect from the breakdown of scale-free localization}

\author{Kazuma Saito}
\affiliation{Department of Applied Physics, Tokyo University of Science, Katsushika, Tokyo 125-8585, Japan}
\affiliation{Institute of Physics, Academia Sinica, Taipei 115201, Taiwan}

\author{Ryo Okugawa}
\affiliation{Department of Applied Physics, Tokyo University of Science, Katsushika, Tokyo 125-8585, Japan}

\author{Kazuki Yokomizo}
\affiliation{Department of Physics, The University of Tokyo, 7-3-1 Hongo, Bunkyo-ku, Tokyo 113-0033, Japan}

\author{Takami Tohyama}
\affiliation{Department of Applied Physics, Tokyo University of Science, Katsushika, Tokyo 125-8585, Japan}

\author{Chen-Hsuan Hsu}
\affiliation{Institute of Physics, Academia Sinica, Taipei 115201, Taiwan}
\affiliation{Physics Division, National Center for Theoretical Sciences, Taipei 106319, Taiwan}

\date{\today}

\begin{abstract}
Non-reciprocal systems exhibit extreme sensitivity to boundary conditions, typically manifesting as the non-Hermitian skin effect (NHSE) under open boundaries. By bridging the boundaries with a tunable impurity bond, one can access intermediate regimes where scale-free localization (SFL) can emerge.
Here, we investigate the competition between such boundary coupling and quasiperiodic disorder in a one-dimensional non-reciprocal lattice.
Our analyses reveal a quasiperiodicity-induced breakdown of the SFL regime, which evolves into either the NHSE or an extended regime, depending on boundary conditions.
These results uncover the crucial roles of boundary effects and quasiperiodicity in non-Hermitian systems.
\end{abstract}

\maketitle

\noindent
{\bf \large{Introduction}}\\
\noindent
In non-Hermitian systems~\cite{Bender2007,Okuma2023}, the non-Hermitian skin effect (NHSE) is one of the most striking phenomena characterized by the exponential localization of a macroscopic number of eigenstates at the system boundaries.
This effect is a hallmark of the Hatano-Nelson model~\cite{Hatano1996,Hatano1997,Hatano1998}, a paradigm lattice system featuring non-reciprocal hoppings.
In this model, complex-valued energies and extended states appear under periodic boundary conditions (PBC), whereas real-valued energies and macroscopic boundary-accumulated states emerge under open boundary conditions (OBC), indicating an extreme sensitivity to boundary conditions.
The NHSE under OBC is topologically rooted in the nontrivial point-gap topology of the energy spectrum under PBC, a phenomenon unique to non-Hermitian systems~\cite{Yao2018,Gong2018,Lee2019,Kawabata2019,Borgnia2020,Okuma2020,Yoshida2020,Zhang2020,Kawabata2020,Okugawa2020,Fu2021,Longhi2021phase,Okugawa2021,Schindler2023,Hamanaka2023,Manna:2023,Nakamura2023,Ma2024,Yoshida2024,Hamanaka2024}.

The introduction of quasiperiodicity into non-Hermitian systems gives rise to a richer class of localization phenomena. In the Hermitian setting, the Aubry-Andre-Harper (AAH) model~\cite{Harper1955,Aubry1980} provides a paradigmatic example of a quasiperiodic system exhibiting a global localization transition of all eigenstates upon tuning the quasiperiodic potential strength. In contrast, in the non-Hermitian extension of the AAH model, increasing the quasiperiodic potential strength leads to distinct boundary-dependent transitions: an NHSE-localization transition under OBC and a delocalization-localization transition under PBC~\cite{Jiang2019,Longhi2019topological,Longhi2019metal,Liu2020nonHermitian,Zeng2020,Liu2020generalized,Liu2021exactnon,Longhi2021nonHermitian,Zhai2022noneq,Chen2022,Zhai2022Kibble,Acharya2024,Zhou2024,Li2024,Padhi2024,Tong2025,Wang2025quasiperiodicity}.

To physically control boundary sensitivity, one can introduce a tunable impurity bond connecting the boundaries. Tuning the hopping strength of this single bond allows for a continuous interpolation between the OBC and PBC limits, effectively imposing generalized boundary conditions (GBC)~\cite{Koch2020,Li2021,Guo2021,Hetenyi:2025a,Wang2025observation}. Crucially, this impurity-induced boundary effect modulates the NHSE into a distinct localized regime known as scale-free localization (SFL)~\cite{Li2020,Li2021,Liu2021exactsol,Yokomizo2021,Molignini2023,Li2023,Wang2023,Jiang2024,Sawada2024,Liu2024,Peng2025,Zhang2025,Shigedomi2025}. In this regime, eigenstates accumulate near the boundaries but possess localization lengths that scale linearly with the system size, contrasting with the size-independent localization characteristic of the NHSE. SFL is therefore a finite-size phenomenon, with its spectral and localization characteristics continuously approaching those of the PBC system in the thermodynamic limit. 
Examining this tunable setting disentangles the roles of boundary effects and system sizes.

Beyond their fundamental properties, the strong boundary-condition sensitivity of non-reciprocal systems inherently leads to numerical instability, demanding high computational precision~\cite{Trefethen2005,Metz2019}. This instability originates from the non-normality of non-Hermitian Hamiltonians, where left and right eigenstates become highly non-orthogonal. Remarkably, the non-normality encodes essential information about the underlying wavefunction properties. The condition number, introduced as a quantitative measure of non-normality and numerical instability~\cite{Trefethen2005,Okuma2020,Ashida2020,Feng2025}, has recently been identified as an indicator for the NHSE~\cite{Nakai2024}. By employing the condition number, one can thus characterize various regimes under GBC, where conventional topological arguments are not directly applicable.

In this work, we investigate the competition between these two distinct localization mechanisms: the quasiperiodicity-induced bulk localization and non-Hermiticity-induced boundary localization, which can manifest as either the NHSE or the SFL and are tunable via a boundary impurity. 
We introduce the non-normality ratio as an efficient indicator capable of distinguishing various regimes and constructing the regime diagram across a broad parameter space.
In contrast to the naive expectation that increasing quasiperiodicity would smoothly drive the SFL into a bulk-localized phase, we report the discovery of a quasiperiodicity-induced NHSE: 
starting from the SFL regime, increasing the quasiperiodic potential strength drives a breakdown of SFL, which evolves into the NHSE regime before entering the bulk-localized phase. Depending on the boundary parameters, an evolution into an extended regime is also observed.
The  breakdown arises from the quasiperiodicity-induced deformation of the GBC spectrum while preserving the point gap of the PBC spectrum. These findings provide insights into the interplay among quasiperiodicity, NHSE, and SFL.
\bigskip

\begin{figure}[t]
    \centering
    \includegraphics[width=0.8\linewidth]{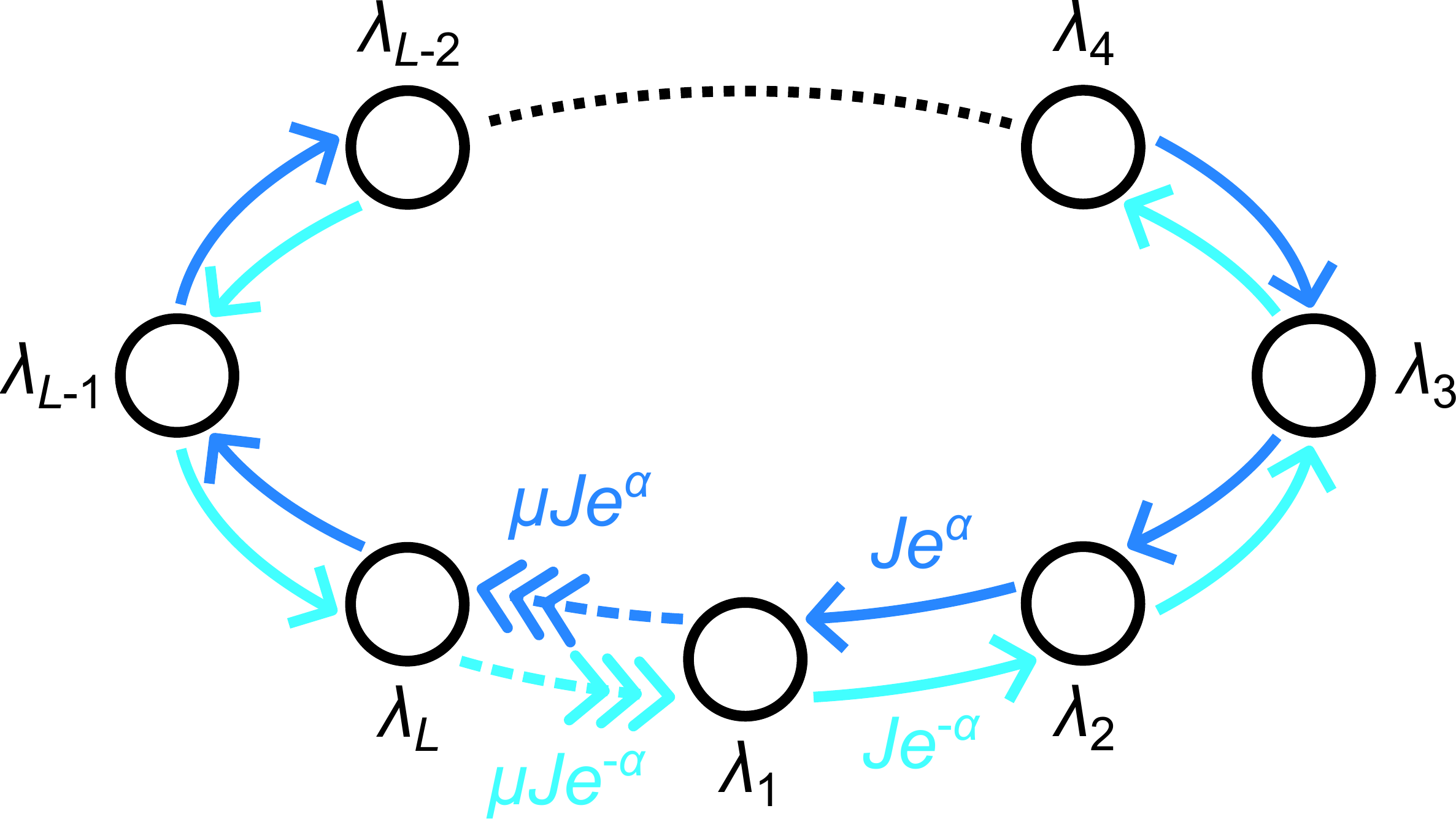}
    \caption{{\bf Non-Hermitian quasiperiodic chain with an impurity bond.} Schematic illustration of the system, with non-reciprocal hopping strength $J e^{ \pm \alpha}$, quasiperiodic onsite potential $\lambda_{j}$, and an impurity bond with hopping strength tunable by $\mu$.
    }
    \label{fig:system}
\end{figure}

\noindent
{\bf \large{Results}}\\
\noindent
{\bf Model}\\
\noindent
We consider a one-dimensional lattice forming a closed loop in which the endpoints are connected by a tunable impurity bond, incorporating non-reciprocal hoppings between nearest-neighbor sites and an onsite quasiperiodic potential, as illustrated in Fig.~\ref{fig:system},
\begin{align}
    H
    &=
    -J \sum_{j = 1}^{L - 1} \ab(e^{\alpha} c_{j}^\dagger c_{j + 1} + e^{-\alpha} c_{j + 1}^\dagger c_{j})
    + \sum_{j = 1}^{L} \lambda_{j} c_{j}^\dagger c_{j}
    \nonumber \\
    &\quad
    - \mu J \ab(e^{\alpha} c_{L}^\dagger c_{1} + e^{-\alpha} c_{1}^\dagger c_{L}).
    \label{eq:H}
\end{align}
Here, $L$ is the total number of sites, $J$ is the hopping amplitude, and $\alpha ~(>0)$ represents the degree of non-reciprocity. The operators $c_{j}^\dagger~(c_{j})$ denote the creation (annihilation) of a fermionic particle at site $j$.
The term $\lambda_{j} \coloneq \lambda \cos(2\pi j / \tau + \phi)$ represents the quasiperiodic potential, where $\lambda$ denotes the potential strength controlling bulk localization, $\tau$ is the wavelength set to the golden ratio, and $\phi$ is a real phase parameter.  
The parameter $\mu$ modifies the hopping amplitude across the impurity bond, effectively tuning the boundary condition. Specifically, $\mu = 0$ corresponds to OBC, while $\mu = 1$ recovers PBC, meaning that a value of $\mu \in (0, 1)$ represents GBC~\cite{Koch2020}. 

The system under GBC exhibits two distinct regimes that can be distinguished by the scaling of the localization length $\xi$.
Assuming that the wavefunction envelope centered at $X$ can be characterized by an exponential profile,
\begin{equation}
    |\psi(x)| \propto e^{-|x-X|/\xi},
    \label{eq:psi_single-center}
\end{equation}
the distinction between the NHSE and SFL regimes is determined by the dependence of $\xi$ on the system size $L$.
In the NHSE regime, $\xi$ remains independent of $L$, whereas in the SFL regime, $\xi$ grows linearly with $L$.

For the present model with $\lambda = 0$, the NHSE regime occurs for $|\ln\mu|>\alpha L$, whereas the SFL regime occurs for $|\ln\mu|<\alpha L$.
Beyond the distinct $L$ dependence, the localization length also depends on $\alpha$ and $\mu$ in general.
Specifically, under OBC, the system lies in the NHSE regime and exhibits $\xi\simeq \alpha^{-1}$.
In contrast, under GBC in the SFL regime, the localization length scales as $\xi\simeq |\ln\mu|^{-1} L$.
Consequently, wavefunctions in the SFL regime remain sensitive to the boundary hopping amplitude, whereas those in the NHSE regime are effectively insensitive to it.
Further details on the regimes for $\lambda = 0$ are provided in Supplementary Note 1.

When quasiperiodicity is introduced ($\lambda\neq0$), bulk localization emerges and competes with the NHSE and SFL phenomena. 
In non-reciprocal systems, the bulk localization transition occurs at the critical strength $\lambda_{\mathrm{c}} = 2J e^{\alpha}$, independent of boundary conditions. Under OBC this critical value marks the NHSE-bulk localization transition~\cite{Jiang2019}, while under PBC it corresponds to the delocalization-localization transition~\cite{Jiang2019,Wang2025quasiperiodicity}.

In the numerical calculations, we set $J$ as the overall energy scale and fix the non-reciprocity parameter at $\alpha = 0.5$ (yielding a critical potential strength $\lambda_{\mathrm{c}} \approx 3.3J$). We vary $\mu$ and $\lambda$ over non-negative values to explore various regimes. While $L$ is generally varied to extract scaling behavior, a Fibonacci number is chosen for fixed-$L$ plots to minimize the mismatch between the lattice sites and the quasiperiodic modulation.
Unless otherwise noted, the quantities are obtained as averages taken over 100 values of $\phi$. With the coexisting quasiperiodicity, non-reciprocity, and an impurity bond, we examine the wavefunction profile $|\psi(x)|$ of the right eigenstate with the second smallest real part, since the smallest one is dominated by the impurity-induced mode and offers limited information about the quasiperiodicity-induced localization.

\noindent
{\bf Non-normality ratio}\\
\noindent
The tunable hopping strength across the impurity bond effectively controls the boundary condition, enabling the realization of GBC.
In the presence of non-reciprocal hopping, such intermediate boundary conditions give rise to the SFL regime~\cite{Li2021,Zhang2025}, where a macroscopic number of states become exponentially localized near the boundaries, with a localization length proportional to the system size.
Since SFL occurs in a finite chain under GBC,
the condition number, which measures the non-normality of the Hamiltonian, provides a useful diagnostic for distinguishing SFL from the NHSE beyond conventional topological characterization.

The condition number is a quantity defined for a diagonalizable Hamiltonian $H$.
Let $V$ be an invertible matrix formed by arranging the right eigenvectors $\ket|E_{\mathfrak{a}}^{a}>$ with eigenvalue $E_{\mathfrak{a}}$ of $H$,
\begin{equation}
    V = \ab(\mqty{\ket*|E_{1}^{1}> & \dots & \ket*|E_{1}^{d_{1}}> & \ket*|E_{2}^{1}> & \dots & \ket*|E_{2}^{d_{2}}> & \dots}),
    \label{eq:def of V}
\end{equation}
which diagonalizes $H$. 
We choose the right eigenvectors to satisfy the biorthonormality condition with the corresponding left eigenvectors $\lket{E_{\mathfrak{a}}^{a}}$, namely,
\begin{equation}
	\lrbraket{E_{\mathfrak{a}}^{a}}{E_{\mathfrak{b}}^{b}} = \delta_{\mathfrak{a}\mathfrak{b}} \delta_{ab}.
	\label{eq:biorthogonal cond.}
\end{equation}
Here, the index $a \in \{ 1, \dots, d_{\mathfrak{a}} \}$, with $d_{\mathfrak{a}}$ denoting the degree of degeneracy. For this matrix $V$, the condition number $\kappa(V)$ is defined as
\begin{equation}
    \kappa = \norm{V} \norm{V^{-1}},
    \label{eq:kappa_V}
\end{equation}
where $\norm{V}$ is the 2-norm of $V$, which is equal to the maximum singular value of $V$.

Since the condition number in Eq.~\eqref{eq:kappa_V} depends on the choice of linear combinations of the degenerate eigenvectors, it is generally not uniquely determined. 
To define a unique condition number, we impose the normalization condition \cite{Nakai2024}
\begin{equation}
    \braket<E_{\mathfrak{a}}^{a}|E_{\mathfrak{a}}^{b}> = \delta_{ab}.
    \label{eq:norm. cond. of right eigenstates}
\end{equation}
In this case, the condition number is equal to the unity if and only if the Hamiltonian is normal.
As analytically derived in Methods section, the condition number scales exponentially with the system size in the NHSE regime, while it obeys a distinct scaling law in the SFL regime.

To effectively distinguish different localization regimes, we introduce the
{\it non-normality ratio} as 
\begin{equation}
\kappa_{\mathrm{R}} (\mu) \coloneq \frac{\kappa_{\mathrm{GBC}} (\mu) }{\kappa_{\mathrm{PBC}}},
\label{eq:kappa_R}
\end{equation} 
where $\kappa_{\mathrm{GBC}}(\mu)$ denotes the condition number of the Hamiltonian under GBC
with impurity-bond strength $\mu$, and $\kappa_{\mathrm{PBC}}$ denotes the one  under PBC with $\mu=1$. 
By normalizing the GBC condition number by its PBC counterpart, bulk contributions are reduced. 
The non-normality ratio $\kappa_{\mathrm{R}}$ thus captures boundary-dependent localization effects.
In the NHSE regime, where the localization length $\xi$ is independent of the system size $L$,
$\kappa_{\mathrm{R}}$ grows exponentially with the leading factor $e^{L/\xi}$,
which originates from the boundary localization.
In the extended or bulk-localized phase, $\kappa_{\mathrm{R}}$ remains $O(1)$ as $L$ varies since the contribution from the boundary effects is absent.
The SFL regime also does not exhibit exponential growth with $L$.
Since its localization length scales linearly with the system size, $\xi=\gamma L$ with $\gamma >0$ being independent of $L$,
the exponential factor $e^{L/\xi}$, which governs the growth of the condition number in the NHSE regime,
becomes $e^{1/\gamma}$ in the SFL regime (see Methods Section).
Nevertheless, 
since $ \gamma$ can depend on the boundary condition as discussed above,
the SFL regime can be characterized by a strong boundary-condition dependence of $\kappa_{\mathrm{R}}$ through $\mu$, as will be discussed below.

\begin{figure}[t]
    \centering 
    \includegraphics[width=\linewidth]{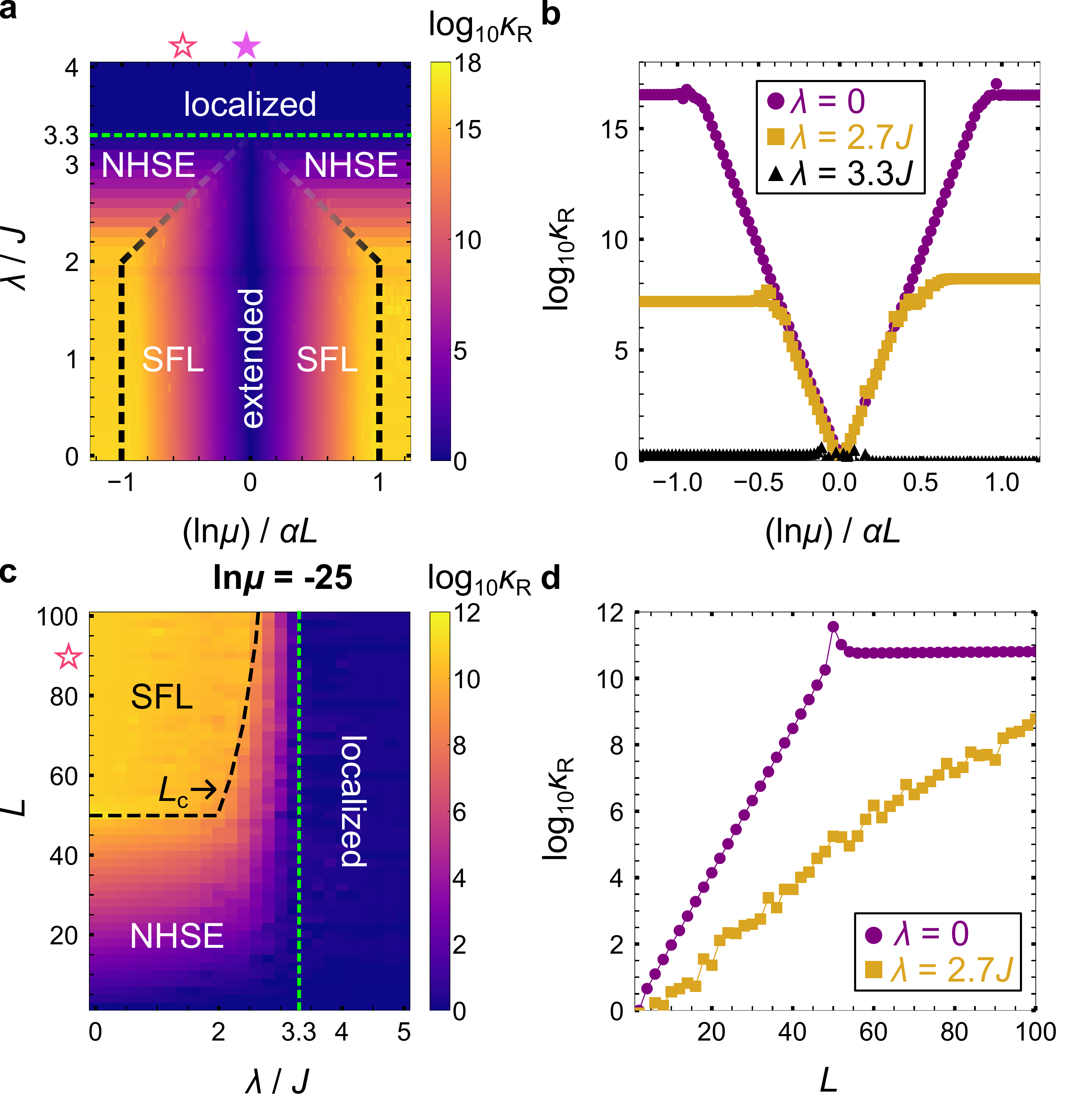}
    \caption{
    {\bf NHSE-SFL regime diagrams.} {\bf a} Non-normality ratio-based regime diagram in the $(\ln\mu,\lambda)$ plane for $L = 89$ and $\phi = 0$. The black-gray dashed line represents the NHSE-SFL boundary. Specific values on the $\ln \mu / (\alpha L)$ axis are indicated by an open star ($\approx -0.56$) and a solid star ($\approx -0.01$). The corresponding parameters for $L = 89$ in {\bf c} and Fig.~\ref{fig:extended}{\bf a} are indicated by the same symbols.
    {\bf b} $\mu$-dependence of $\log_{10}\kappa_{\mathrm{R}}$ corresponding to fixed-$\lambda$ cuts in {\bf a}. 
    {\bf c} A regime diagram on the $(\lambda,L)$ plane for $\ln \mu = -25$. The dashed curves correspond to the boundary shown in {\bf a}.
    {\bf d} Scaling behavior of $\kappa_{\mathrm{R}}$ for the parameters used in {\bf c}.
    }
    \label{fig:kappa}
\end{figure}

\noindent
{\bf Regime diagram for general $\mu$ and $\lambda$}\\
\noindent
Figure~\ref{fig:kappa}{\bf a} displays the regime diagram in the $(\ln \mu, \lambda)$ plane,
determined from the behavior of the condition number and wavefunctions.
The horizontal axis is rescaled by $\alpha L$, where the values of $\alpha$ and $L$ are fixed.
Our numerical verification confirms that the diagram is obtained independently of $L$ for sufficiently large system sizes.
Here, the line of $\ln\mu = 0$ corresponds to the PBC limit, where the system undergoes a sharp transition from the extended to the localized phase at $\lambda_{\mathrm{c}}$.
The green dashed line indicates this critical $\lambda_{\mathrm{c}}$ value, representing the bulk localization transition.
This transition persists for any $\mu$ because it is characterized by the behavior of bulk wavefunctions.
The diagram is consistent with the one  obtained from entanglement spectra; see Supplementary Note 2 for details.
Although we focus on the region $\ln \mu <0$ here, the corresponding results for $\ln \mu > 0$ and the regime diagram for general $\mu$ are shown in Supplementary Note 3.

Interesting features emerge in the region $\lambda<\lambda_{\mathrm{c}}$ for $0<|\ln\mu|<\alpha L$, where the scaling of the non-normality ratio generally depends on the potential strength $\lambda$ and the boundary condition set by $\mu$.
To examine these features more closely,  in Fig.~\ref{fig:kappa}{\bf b} we show the dependence of the non-normality ratio on the boundary hopping amplitude for fixed $\lambda$. 
For $\lambda<\lambda_{\mathrm{c}}$, the dependence of $\log_{10}\kappa_{\mathrm{R}}$ on $|\ln\mu|$ separates into two regimes: a linearly increasing region and a plateau where the value remains nearly constant.
Since $\alpha$ and $L$ are fixed, the plateau indicates that the wavefunctions are insensitive to the boundary hopping amplitude.
These broad plateau regions for $\lambda < \lambda_c$ correspond to the NHSE regime, as confirmed by the scaling behavior of $\kappa_{\mathrm{R}}$ with $L$.
Meanwhile, the linear region appears between the NHSE and extended regimes.
In this linear region, $\kappa_{\mathrm{R}}$ depends strongly on $\mu$, reflecting the sensitivity of the wavefunctions to the boundary hopping amplitude. This behavior is characteristic of the SFL regime, as discussed above.

Notably, the value of $\mu$ at which the crossover between the linear and plateau regions occurs depends on $\lambda$, as indicated by the black-gray dashed line in the $(\ln\mu,\lambda)$ plane in Fig.~\ref{fig:kappa}{\bf a}.
For $\lambda=0$, the numerical results in Fig.~\ref{fig:kappa}{\bf b} show that $\log_{10}\kappa_{\mathrm{R}}$ is proportional to $|\ln\mu|$, consistent with the analytical estimation of $\gamma \simeq |\ln\mu|^{-1}$.
As $\lambda$ approaches $\lambda_{\mathrm{c}}$, $\kappa_{\mathrm{R}}$ tends toward unity because the system approaches the bulk-localized regime.
Consequently, for $\lambda\lesssim\lambda_{\mathrm{c}}$, the SFL region is compressed, making the regime identification less clear. This behavior is reflected in the color gradient of the dashed crossover line.
Overall, the change in the $\mu$ dependence of $\kappa_{\mathrm{R}}$ allows us to identify a crossover between the SFL and NHSE regimes.
 
In Figs.~\ref{fig:kappa}{\bf c} and \ref{fig:kappa}{\bf d}, we present the scaling behavior of the non-normality ratio for $\mu\ll1$.
Figure~\ref{fig:kappa}{\bf c} shows the results in the $(\lambda,L)$ plane under GBC with $\ln\mu=-25$, where an NHSE-SFL crossover occurs at a $\lambda$-dependent critical length $L_{\mathrm{c}}$ (black curve) within the regime $\lambda<\lambda_{\mathrm{c}}$.
In the NHSE regime, $\kappa_{\mathrm{R}}$ increases exponentially with $L$, as shown in Fig.~\ref{fig:kappa}{\bf d}.
For $L>L_{\rm c}$, the scaling crosses over to an approximately logarithmic behavior.
At $\lambda=0$, $\kappa_{\mathrm{R}}$ reaches the order of $10^{10}$ and the critical length is $L_{\rm c}=50$, consistent with the analytical estimate $L_{\rm c}=|\ln\mu|/\alpha$ presented in Supplementary Note 1 and in previous studies~\cite{Li2021,Wang2023}.
For nonzero $\lambda$, we obtain the critical length from fitting the crossover line between the NHSE and SFL regimes for 
$\lambda > 2J$  in Fig.~\ref{fig:kappa}{\bf a}.
The result roughly follows $L_{\mathrm{c}}\propto1/|\lambda-\lambda_{\rm c}|$ for $\lambda > 2J$, corresponding to the crossover curve shown in Fig.~\ref{fig:kappa}{\bf c}.
For example, at $\lambda=2.7J$, we estimate $L_{\mathrm{c}}=109$.
Therefore, when $\mu\ll1$, increasing the quasiperiodic potential can drive a crossover from the SFL regime to the NHSE regime in sufficiently large finite systems.

\begin{figure}[t]
    \centering
    \includegraphics[width=\linewidth]{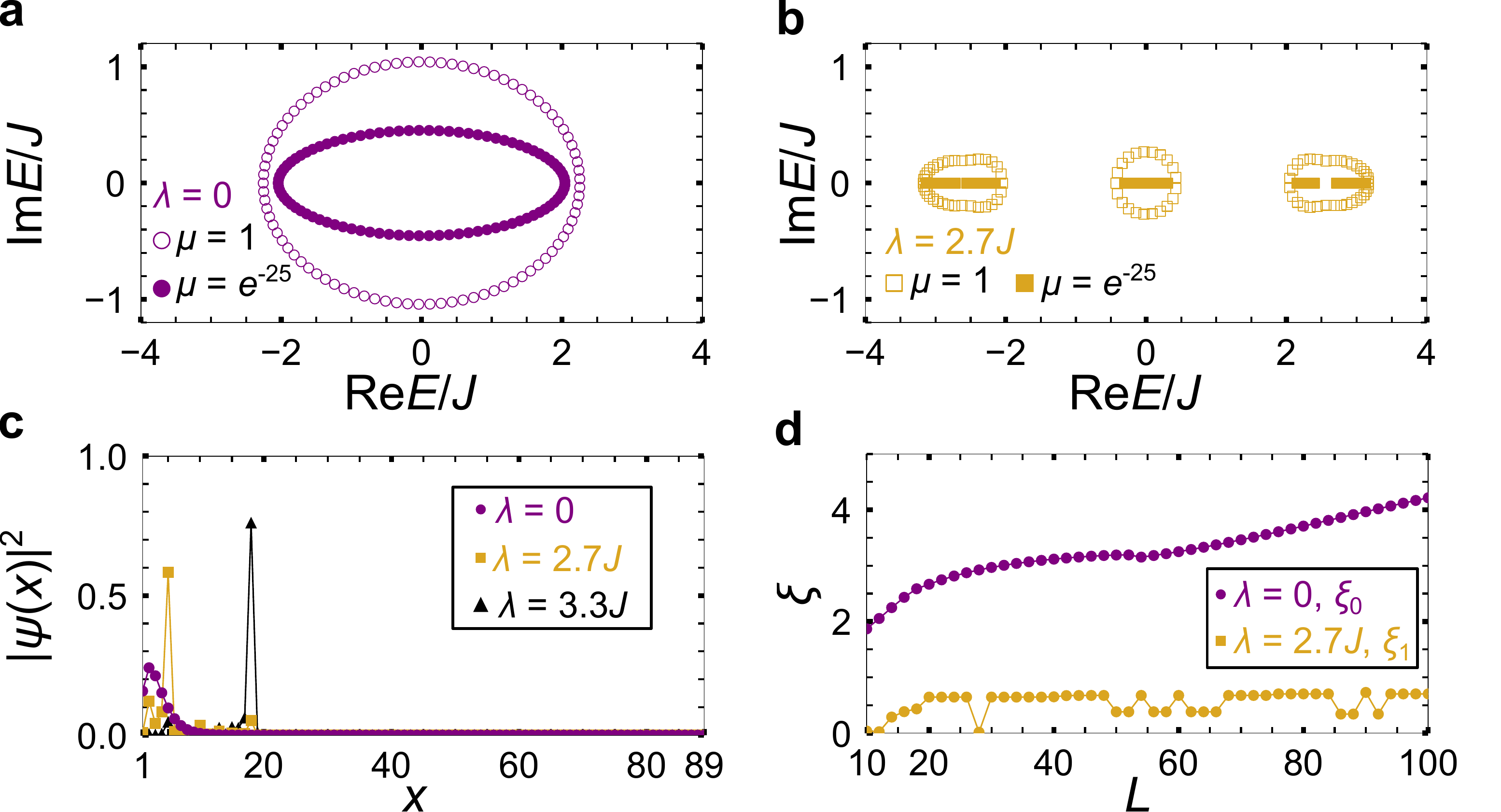}
    \caption{{\bf Spectral behaviors in the localization regimes. } Characteristics of various regimes for $\phi = 0$  and representative $\lambda$ and $\mu$ values.
    {\bf a},{\bf b} Energy spectra for $L = 89$. The closed (open) symbols represent the spectra for $\ln \mu = {-25}$ ($\mu = 1$).
    {\bf c} Spatial distributions of the density $|\psi(x)|^2$ for $L = 89$ and $\ln \mu = {-25} $.
    {\bf d} Localization lengths, $\xi_{0}$ (for $\lambda = 0$) and $\xi_{1}$ (for $\lambda = 2.7 J$), for $\ln \mu = {-25}$.  
    }
    \label{fig:NHSE}
\end{figure}

\noindent
{\bf Quasiperiodicity-induced NHSE} \\
\noindent
To examine the NHSE regime induced by sufficiently strong $\lambda$, we analyze the energy spectrum, the spatial distribution of eigenstates, and the localization length for representative $\mu$ and $\lambda$ values, as displayed in Fig.~\ref{fig:NHSE}.
We first examine the PBC ($\ln \mu = 0$) and GBC $(\ln \mu = -25)$ energy spectra for a fixed $L$. For $\lambda = 0$, as shown in Fig.~\ref{fig:NHSE}{\bf a}, the GBC spectrum remains complex and lies inside the PBC one as typical of the SFL regime~\cite{Li2020,Li2021,Liu2021exactsol,Yokomizo2021,Molignini2023,Li2023,Wang2023,Liu2024,Peng2025,Zhang2025,Shigedomi2025}.
In stark contrast, for $\lambda = 2.7J$, the GBC spectrum completely collapses onto the real axis within the complex PBC loop, as shown in Fig.~\ref{fig:NHSE}{\bf b}. This spectral feature provides unambiguous evidence of the genuine emergence of the NHSE regime, marking a breakdown of the SFL regime observed in the absence of quasiperiodicity.

Figure~\ref{fig:NHSE}{\bf c} shows the spatial distribution of the density $|\psi(x)|^2$.
We identify distinct peak positions characteristic of the SFL ($\lambda = 0$), NHSE ($\lambda = 2.7J$), and localized ($\lambda = 3.3J$) regimes for a fixed $L$, denoted as $X_{0}$, $X_1$, and $X_2$, respectively.
In the intermediate range $2J \lesssim \lambda \lesssim 3.3J$, peaks at these positions coexist; as $\lambda$ increases, the intensity at $X_0$ continuously diminishes while the one at $X_1$ becomes dominant, followed by the gradual emergence of the bulk peak at $X_2$. This continuous transfer of wavefunction weight between spatially distinct modes justifies the identification of the SFL-NHSE boundary as a crossover.
To quantitatively analyze this behavior and extract the corresponding localization lengths, we model the wavefunction profile of (right) eigenstate with the second-smallest real part of its eigenvalue using a multi-peak ansatz with the peak positions $X_{0, 1, 2}$:
\begin{equation}
    |\psi(x)| = A_0 e^{-|x - X_0| / \xi_0} + A_1 e^{-|x - X_1| / \xi_1} + A_2 e^{-|x - X_2| / \xi_2},
    \label{eq:fitting func. for wave func.}
\end{equation}
where $A_{0, 1, 2}$ and $\xi_{0, 1, 2}$ are the fitting parameters representing the amplitudes and localization lengths of the respective components.

Figure~\ref{fig:NHSE}{\bf d} shows the scaling of the localization lengths $\xi_{0}$ and $\xi_{1}$ obtained by fitting the wavefunction to Eq.~\eqref{eq:fitting func. for wave func.}.
For $\lambda = 0$, only the peak at $X_0$ exists. Except for very small systems,
$\xi_0$ changes from being approximately constant to scaling linearly with $L$ beyond a critical system size, indicating a crossover from the NHSE regime to the SFL regime around the critical length $L_{\mathrm{c}} = 50$~\cite{Li2021}.
In contrast, for $\lambda = 2.7 J$, peaks at $X_0$, $X_1$, and $X_2$ are present, with a dominant peak at $X_1$. The value of $\xi_1$ is nearly independent of $L$, exhibiting NHSE behavior, even beyond $L = 50$.
We find an NHSE regime appearing between $\lambda = 2J$ (the critical value in the Hermitian limit, $\alpha = 0$) and the corresponding critical value, $\lambda_{\mathrm{c}} \approx 3.3J$, of the present system, as shown in Fig.~\ref{fig:kappa}{\bf a}.
By contrast, in systems with purely unidirectional hopping~\cite{Zhang2025}, the transition between the SFL regime and the localized phase occurs exactly at $\lambda = 2J$, and this quasiperiodicity-induced NHSE regime is absent in the large-$L$ limit, as it shrinks to a negligible range.

Due to the effect of quasiperiodicity, the spectral loops close as the system approaches the bulk localization transition.
Since the GBC spectrum lies inside the PBC spectrum, it collapses into line segments at a smaller $\lambda$ value than the PBC spectrum.
Consequently, an NHSE regime emerges between the SFL regime and the bulk-localized phase.
From a real-space perspective, the effective OBC is realized because the quasiperiodic potential suppresses the hopping across the impurity bond prior to the hoppings in the bulk; further details are provided in Supplementary Note 4.
In the thermodynamic limit, the emerging NHSE region collapses to the critical point $\lambda_{\mathrm{c}}$, where the SFL regime evolves into the extended regime as $\xi \to \infty$. Consequently, for $L \to \infty$, the system exhibits a delocalization-localization transition analogous to that under PBC.

 \begin{figure}[t]
    \centering
    \includegraphics[width=\linewidth]{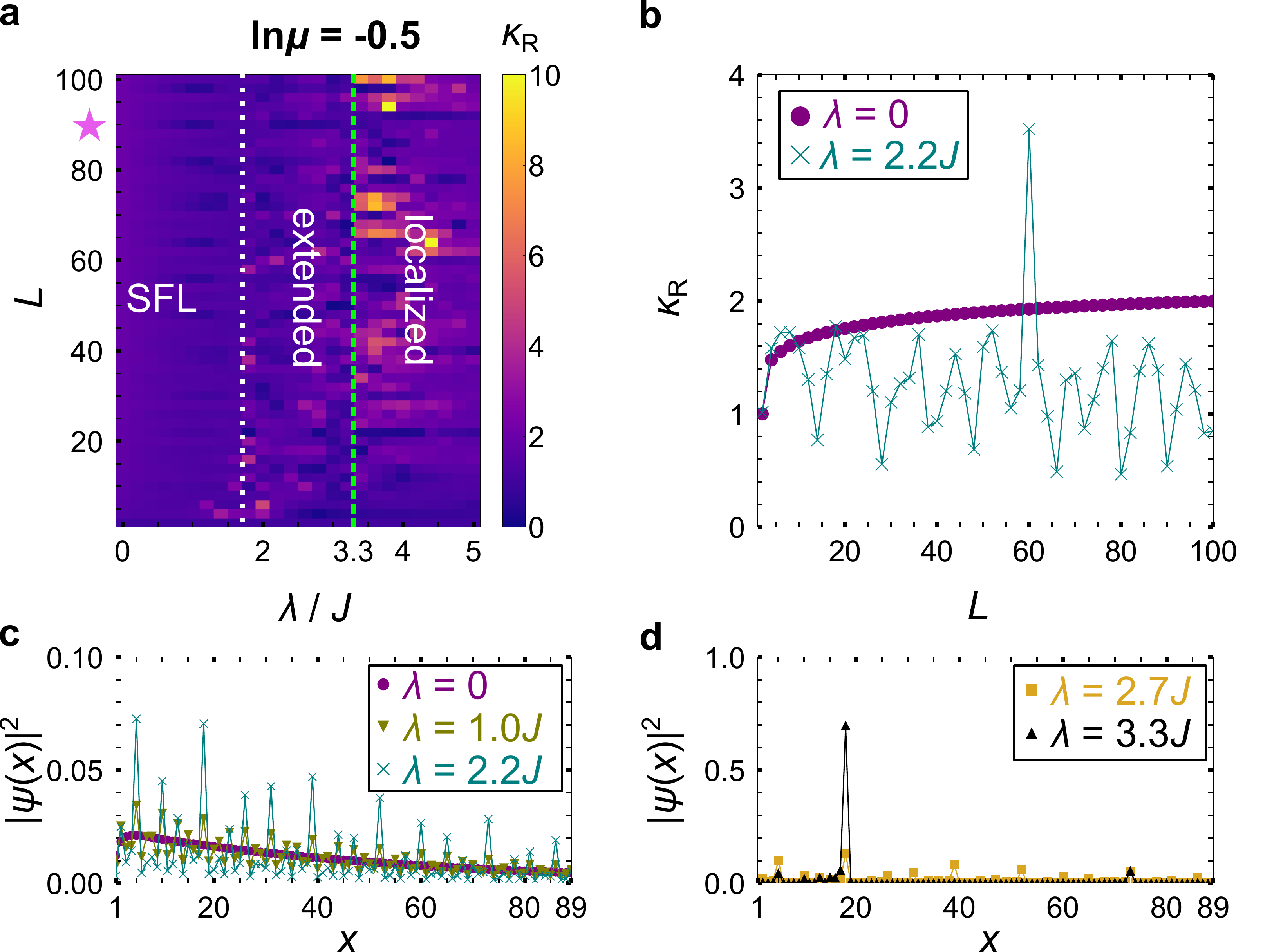}
    \caption{{\bf Regimes in the proximity to the PBC limit.}
    {\bf a} A regime diagram on the $(\lambda,L)$ plane for $\ln \mu = -0.5$. The dashed vertical curves correspond to the boundary shown in Fig.~\ref{fig:kappa}{\bf a}. The solid star at $L = 89$ corresponds to the same symbol in Fig.~\ref{fig:kappa}{\bf a} and represents $\ln\mu / (\alpha L) \approx -0.01$.
    {\bf b} Scaling behavior of $\kappa_{\mathrm{R}}$ for the parameters used in {\bf a}.
    {\bf c, d} Spatial distributions of the density $|\psi(x)|^2$ for $L = 89$, $\ln \mu = {-0.5}$ and  $\phi = 0$ for SFL and extended states ({\bf c}) and for extended and localized states ({\bf d}).
     }
    \label{fig:extended}
\end{figure}

\noindent
{\bf Quasiperiodicity-assisted delocalization.} \\
\noindent
Figure~\ref{fig:extended}{\bf a} shows the non-normality ratio in the region near the PBC limit with $\ln\mu = -0.5$.
Here, the system exhibits the SFL regime for small $\lambda$.
At $\lambda = 0$, the numerical results show that $\kappa_{\mathrm{R}}$ increases logarithmically with $L$, as shown in Fig.~\ref{fig:extended}{\bf b}.
As the potential strength increases within the range $\lambda < 1.7 J$, $\kappa_{\mathrm{R}}$ shows increasing behavior while the growth rate is gradually suppressed.
This increasing trend vanishes at $\lambda \approx 1.7 J$, which implies that the localization properties of the wavefunction begin to coincide with those under PBC.
Therefore, our results suggest that when the system is under GBC but remains close to the PBC limit, increasing $\lambda$  drives a crossover from the SFL regime to an extended regime.

As illustrated in Fig.~\ref{fig:extended}{\bf c}, the SFL profile gradually evolves into an extended wave-like form as $\lambda$ increases.
The wavefunction analysis clarifies that the region $(1.7J<\lambda<\lambda_{\mathrm c})$ is an extended regime.
Upon further increasing $\lambda$,
the system undergoes a delocalization-localization transition at $\lambda = \lambda _c$, similar to that under PBC, as shown in Fig.~\ref{fig:extended}{\bf d}.
In this regime, the PBC ($\ln \mu = 0$) and GBC ($\ln \mu = -0.5$) spectra are nearly identical; increasing $\lambda$ suppresses the difference even further for $\lambda < \lambda _c$.
As a result, the system undergoes a crossover from the SFL regime to an extended regime. Because $\mu$ is close to unity, the boundary conditions effectively approach PBC, favoring the extended regime.
This behavior contrasts with the large $|\ln\mu|$ case, where the intermediate-$\lambda$ regime exhibits the NHSE.
Such quasiperiodicity-assisted delocalization underscores another nontrivial interplay between quasiperiodicity and the SFL regime.

\bigskip

\noindent
{\bf \large{Discussion}} \\
\noindent
Generally, increasing the quasiperiodic potential is expected to suppress the NHSE or extended states, driving the system toward bulk localization.
Contrary to this conventional scenario, we find that when the system is in the SFL regime, quasiperiodicity destroys this state, inducing an extended regime near the PBC limit and the NHSE regime far from it, before eventually entering the boundary-independent localized phase.
We attribute the origin of this behavior to the distinct responses of the PBC spectrum and the GBC spectrum to the quasiperiodic potential.
Furthermore, our analysis of the condition number shows that wavefunctions in the SFL regime, as a finite-size phenomenon, are highly sensitive to the boundary hopping amplitude. This behavior contrasts with the NHSE regime, where the condition number is insensitive to the boundary hopping amplitude.

In non-reciprocal systems, characterization of different regimes is typically based on the spectral winding number~\cite{Gong2018,Okuma2020,Claes2021}, a topological invariant defined only under PBC and thus inapplicable under GBC. 
Alternative approaches employ boundary-condition-insensitive quantities, such as the entanglement entropy~\cite{Kawabata2023,Wang2023,Zhou2024,Li2024}, but their numerical evaluation requires high precision and is computationally demanding; for completeness, we present the corresponding entanglement-based regime diagram of our system in Supplementary Note 2. 
We further note that the accuracy of the entanglement entropy itself depends on the condition number~\cite{Feng2025}.
Therefore, characterizing regimes through the introduced non-normality ratio, quantified directly by the condition number under GBC, offers a more direct and efficient approach.
 
Non-reciprocal systems with non-Hermitian impurities and random disorder have already been experimentally realized in electrical circuits~\cite{Wang2025observation}, where the SFL has also been observed.
Replacing random disorder with quasiperiodic modulation offers a natural route toward realizing the present model in experiments.
Such a setup would enable direct observation of the predicted quasiperiodicity-assisted NHSE and delocalization, providing a versatile platform to explore the interplay among quasiperiodicity, non-Hermiticity, and boundary conditions.
\bigskip

\noindent
{\bf \large{Methods}}\\
\noindent 
{\bf Bounds on the condition number in the NHSE and SFL regimes}\\
\noindent 
We clarify the behavior of the condition number in the NHSE and SFL regimes.
For a Hamiltonian $H$ with the matrix dimension $L$,
the condition number $\kappa$ in Eq.~\eqref{eq:kappa_V} satisfies the following inequality \cite{Nakai2024}:
\begin{equation}
	\sqrt{\frac{1}{L}\sum_{\mathfrak{a}, a} \llbraket*{E_{\mathfrak{a}}^{a}}{E_{\mathfrak{a}}^{a}}} \leq \kappa \leq \sqrt{L\sum_{\mathfrak{a}, a} \llbraket*{E_{\mathfrak{a}}^{a}}{E_{\mathfrak{a}}^{a}}}.
	\label{eq:kappa bounds of lbraket}
\end{equation}
Here, the left and right eigenvectors are chosen to satisfy the biorthonormality condition given in Eqs.~\eqref{eq:biorthogonal cond.} and \eqref{eq:norm. cond. of right eigenstates}. 
Thus, $\llbraket*{E_{\mathfrak{a}}^{a}}{E_{\mathfrak{a}}^{a}}$ determines the scale of the bounds on the condition number.

Using the inequality in Eq.~\eqref{eq:kappa bounds of lbraket}, we evaluate bounds on the condition number in the NHSE and SFL regimes for a one-dimensional chain of length $L ~(\gg 1)$ under GBC. 
To describe the two regimes,
we assume that almost all right and left eigenvectors can be represented as
\begin{equation}
    \psi_{\mathrm{R}}^{n}(j) = A_{\mathrm{R}}^{n} f^{n}_j(\mu, L)e^{-j / \tilde{\xi}_{n}},~~\psi_{\mathrm{L}}^{n}(j) = A_{\mathrm{L}}^{n} g^{n}_j(\mu, L)e^{j / \tilde{\xi}_{n}},
	\label{eq:right and left eigenfunctions}
\end{equation}
where $n=(\mathfrak{a}, a)$ labels eigenstates, $j=1,\dots,L$ denotes the site index,
$A_{\mathrm{R}(\mathrm{L})}^{n}$ is the normalization constant for the right (left) eigenvector $\psi_{\mathrm{R}(\mathrm{L})}^{n}$, and $\tilde{\xi} _n=s_n\xi _n$ where $\xi _n $ is the localization length and $s_n=\pm 1$ specifies the localization direction. 
Under PBC, since localization is absent, we take $\xi _n\rightarrow \infty$ last. 
Here, the functions $f^n_j(\mu, L)$ and $g^n_j(\mu, L)$ represent modulation factors that are subexponential in $L$ uniformly for $1\le j\le L$,
so that the leading exponential localization behavior is encoded in the factors $e^{\mp j/\tilde{\xi}_n}$.
In the following, we consider a model without internal degrees of freedom for simplicity.

Under the normalization conditions in Eqs.~\eqref{eq:biorthogonal cond.} and \eqref{eq:norm. cond. of right eigenstates}, we obtain
\begin{align}
	\eta_{n} \coloneq \llbraket*{E_{\mathfrak{a}}^{a}}{E_{\mathfrak{a}}^{a}} &= \frac{S_{f^{n}}(\mu, L) S^{\prime}_{g^{n}}(\mu, L)}{|B_{n}(\mu, L)|^{2}} 
	\label{eq:eta_n} \\
	S_{f^{n}}(\mu, L) &\coloneq \sum_{j = 1}^{L} e^{-2j / \tilde{\xi}_{n}} |f^{n}_j(\mu, L)|^{2},
	\label{eq:S_f_L} \\
	S^{\prime}_{g^{n}}(\mu, L) &\coloneq \sum_{j = 1}^{L} e^{2j / \tilde{\xi}_{n}} |g^{n}_j(\mu, L)|^{2},
	\label{eq:Sprime_g_L} \\
    B_n(\mu, L)&\coloneq \sum _{j=1}^Lg^{n*}_j(\mu,L)f^n_j(\mu,L).
\end{align}
To estimate $\eta _n$, we assume that $|B_{n}(\mu, L)|^{-2}$ does not yield an exponential contribution in $L$.
We also assume that $f^{n}_j(\mu, L)$ and $g^{n}_j(\mu, L)$ are nonzero at least at one site near the boundaries where the corresponding mode is localized.
For definiteness, we take them to be nonzero at the boundary site, $j=1$ or $j=L$, depending on the localization direction.

We evaluate the upper bound on $\eta _n$. Using the maximum values of the modulation factors, we find 
\begin{align}
	\eta _n& \leq 
    C_n(\mu, L)\sum_{j = 1}^{L} e^{-2j / \xi_{n}}\sum_{j' = 1}^{L} e^{2j' / \xi_{n}} \\
    &= C_{n}(\mu, L) 
    \left[ \frac{\sinh (L/\xi _n)}{\sinh (1/\xi _n)} \right] ^2 ,
	\label{eq:upper bound of eta_n}
\end{align}
where 
\begin{equation}
C_n(\mu, L):=\frac{\max_{j} |f^{n}_j(\mu, L)|^{2}\max_{j^{\prime}} |g^{n}_{j'}(\mu, L)|^{2}}{|B_n(\mu, L)|^2}.
\end{equation}

Next, we evaluate a lower bound on $\eta _n$. 
For $s_n=+1$, the right eigenstate is localized near the left boundary, whereas the
corresponding left eigenstate is localized near the right boundary.
Because $f^{n}_1(\mu, L)\neq 0$ and $g^{n}_L(\mu, L) \neq 0$ for $s_n=+1$ by assumption,
we have
\begin{align}
	&S_{f^{n}}(\mu, L) \geq |f^{n}_1(\mu, L)|^{2} e^{-2 / \xi_{n}},
	\label{eq:min. evaluation of S_f_L} \\
	&S_{g^{n}}'(\mu, L) \geq |g^{n}_L(\mu, L)|^{2} e^{2L / \xi_{n}}
	\label{eq:min. evaluation of S_g_L} . 
\end{align}
Combining these inequalities, we obtain
\begin{equation}
	\eta_{n} \geq c_{n}(\mu, L) e^{2(L-1) / \xi_{n}},
	\label{eq:lower bound of eta_n}
\end{equation}
where
\begin{align}
    c_n(\mu, L):=\frac{|f^{n}_1(\mu, L)|^{2}|g^{n}_L(\mu, L)|^{2}}{|B_n(\mu, L)|^2}. 
\end{align}
If $s_n=-1$, the same estimate is obtained by interchanging the two boundaries.
In this case,
we have
\begin{align}
    c_n(\mu, L)=\frac{|f^{n}_L(\mu, L)|^{2}|g^{n}_1(\mu, L)|^{2}}{|B_n(\mu, L)|^2}.
\end{align}

The upper and lower bounds on $\eta_{n}$ are given in Eqs.~\eqref{eq:upper bound of eta_n} and \eqref{eq:lower bound of eta_n}. 
By introducing $\xi _M\coloneq \max _n\xi _n, ~\xi _m\coloneq \min _n \xi _n$,
the inequality for $\eta_{n}$ becomes
\begin{equation}
    c_{n}(\mu, L) e^{2(L-1) / \xi_{M}} \leq \eta_{n} \leq C_{n}(\mu, L) \left[ \frac{\sinh (L/\xi _m)}{\sinh (1/\xi _m)} \right] ^2 ,
	\label{eq:inequality of eta_n}
\end{equation}
where we have used the fact that  ${\sinh (L/\xi )}/{\sinh (1/\xi )}$ is a monotonically decreasing function of $\xi >0$. 
For $L\gg 1$, by substituting this inequality into Eq.~\eqref{eq:kappa bounds of lbraket}, we find 
\begin{align}
	&\frac{e^{L / \xi_{M}}}{\sqrt{L}} \sqrt{\sum_{n} c_{n}(\mu, L)} \lesssim \kappa _{\mathrm{GBC}}(\mu)
    \nonumber \\
    &\lesssim \sqrt{L} \frac{\sinh (L/\xi _m)}{\sinh (1/\xi _m)} \sqrt{\sum_{n} C_{n}(\mu, L)}.
	\label{eq:inequality of kappa_V}
\end{align}
We note that $c_n(L)$ and $C_n(L)$ do not scale exponentially with $L$.

We now estimate the scaling behavior of the non-normality ratio $\kappa _{\mathrm{R}}(\mu)$ in the NHSE and SFL regimes.
Due to $\xi _n\rightarrow \infty$ under PBC in these regimes, we have
\begin{equation}
    \frac{1}{\sqrt{L}} \sqrt{\sum_{n} c_{n}(1, L)} \lesssim \kappa _{\mathrm{PBC}}
    \lesssim L^{3/2} \sqrt{\sum_{n} C_{n}(1, L)}.
\end{equation}
As expected, $\kappa _{\mathrm{PBC}}$ does not scale exponentially in $L$.
In the NHSE regime under GBC, since the localization length is independent of the system size,
both the upper and lower bounds in Eq.~\eqref{eq:inequality of kappa_V} contain exponential factors in $L$.
As a result, the non-normality ratio in the NHSE regime exhibits exponential scaling with $L$, which is consistent with Ref.~\cite{Nakai2024}.
In the SFL regime, the localization length is proportional to the system size.
By defining $\gamma _n:=\xi _n/L $,
we have $[\sinh(1 / \xi_{n}) ]^{-1} \simeq \gamma _nL=O(L)$ for $L\gg 1$.
Thus, in the SFL regime,
we obtain 
\begin{align}
	&\frac{e^{1 / \gamma_{M}}}{\sqrt{L}} \sqrt{\sum_{n} c_{n}(\mu, L)} \lesssim \kappa _{\mathrm{GBC}}(\mu)
    \nonumber \\
    &\lesssim \gamma _m \sinh (1/\gamma _m) L^{3/2}\sqrt{\sum_{n} C_{n}(\mu, L)}.
    \label{eq:inequality of kappa_V_SFL}
\end{align}
Consequently, the non-normality ratio in the SFL regime does not exhibit exponential scaling with $L$.
The exponential scaling of $\kappa_{\mathrm{R}}(\mu)$ with system size provides a diagnostic for distinguishing the NHSE and SFL regimes.

In addition, the upper and lower bounds in Eq.~\eqref{eq:inequality of kappa_V_SFL} contain the characteristic exponential factors $e^{1/\gamma_m}$ and $e^{1/\gamma_M}$, respectively, which originate from SFL.
Thus, the non-normality ratio in the SFL regime is strongly sensitive to parameters that control the localization length.
In our model, this sensitivity can be observed by varying the boundary condition.
This behavior is useful to distinguish the SFL regime from extended and bulk-localized regimes
because the condition number in the latter regimes is typically insensitive to boundary conditions.
\bigskip

\noindent
{\bf \large{Data availability}}\\
\noindent
The data that support the findings of this study are available at Zenodo~\cite{data}.
\bigskip

\noindent
{\bf \large{Code availability}}\\
\noindent
The codes used to calculate the condition number are available from the corresponding author upon request.

\bigskip

\noindent
{\bf \large{Acknowledgments}}\\
\noindent
We thank N.~Okuma, M.~Oshikawa, and K.~Totsuka for interesting discussions.
This work was financially supported by the National Science and Technology Council (NSTC), Taiwan (Grant No.~NSTC-114-2112-M-001-057), Academia Sinica (AS), Taiwan  (Grant No.~AS-iMATE-114-12), JST SPRING, Japan (Grant No.~JP-MJSP2151), and JSPS KAKENHI, Japan  (Grant No.~23K13027, No.~JP23K13033, No.~JP24K00586, and No.~JP25H01248).
\bigskip

\noindent
{\bf \large{Author contributions}}\\
\noindent
K.S. and R.O. developed the method for regime discrimination using the condition number and performed the analytical calculations. K.S. carried out all numerical computations. C.-H.H. supervised the project. All authors discussed the results and contributed to the writing and editing of the manuscript.
\bigskip

\noindent
{\bf \large{Competing interests}}\\
\noindent
The authors declare no competing interests.
\bigskip

\noindent
{\bf \large{Additional information}}

\noindent
{\bf Supplementary information} Supplementary material available at http://XXX.

\end{bibunit}

\clearpage
\begin{widetext}
\begin{bibunit}[apsrev4-2]

\setcounter{equation}{0}
\setcounter{figure}{0}
\setcounter{table}{0}
\renewcommand{\figurename}{Supplementary Fig.}
\renewcommand{\theequation}{S\arabic{equation}}
\renewcommand{\thefigure}{\arabic{figure}}
\renewcommand{\thetable}{S\arabic{table}}

\begin{center}
{\Large\textbf{Supplementary Information for ``Quasiperiodicity-induced non-Hermitian skin effect from the breakdown of scale-free localization''}}

\fontsize{10}{12}
Kazuma Saito,$^{1, 2}$ Ryo Okugawa,$^{1}$ Kazuki Yokomizo,$^{3}$ Takami Tohyama,$^{1}$ and Chen-Hsuan Hsu$^{2,4}$ \\
$^{1}${\it Department of Applied Physics, Tokyo University of Science, Katsushika, Tokyo 125-8585, Japan
} \\ 
$^{2}${\it Institute of Physics, Academia Sinica, Taipei 11529, Taiwan
} \\
$^{3}${\it Department of Physics, The University of Tokyo, 7-3-1 Hongo, Bunkyo-ku, Tokyo 113-0033, Japan
} \\
$^{4}${\it Physics Division, National Center for Theoretical Sciences, Taipei 106319, Taiwan}

\end{center}

\section{Supplementary Note 1: Analysis of the system in the absence of the quasiperiodic potential
\label{Appendix:no-qp}
}
    In this section, we review the basic properties of our system when the quasiperiodic potential is absent. 
By applying the imaginary gauge transformation
\begin{equation}
    c_{j} \rightarrow e^{-\alpha j} c_j, ~~c_{j}^\dagger \rightarrow e^{\alpha j} c_{j}^\dagger,
    \label{Seq:imag. gauge trans.}
\end{equation}
the Hamiltonian~(1) without the quasiperiodic potential can be rewritten as
\begin{equation}
    H' = -J \sum_{j = 1}^{L - 1} \ab(c_{j}^\dagger c_{j + 1} + c_{j + 1}^\dagger c_{j}) - \mu J \ab(e^{\alpha L} c_{L}^\dagger c_{1} + e^{-\alpha L} c_{1}^\dagger c_L).
    \label{Seq:imag. gauge transformed H}
\end{equation} 
The relation between the hopping amplitude across the impurity bond and that within the bulk determines whether the energy eigenvalues under the generalized boundary conditions (GBC) are real or complex~\cite{Guo2021,Wang2023}.

We now discuss crossovers between the NHSE and SFL regimes.
For simplicity, we take $\alpha$ to be positive.
In the regime where $\mu ~(>0)$ is sufficiently small, the hopping across the impurity bond can be approximated to be unidirectional and the problem can be treated analytically.
For this purpose, we consider this approximated model, 
\begin{equation}
    H' \simeq -J \sum_{j = 1}^{L - 1} \ab(c_{j}^\dagger c_{j + 1} + c_{j + 1}^\dagger c_{j}) - \mu J e^{\alpha L} c_{L}^\dagger c_{1},
    \label{Seq:H'}
\end{equation}
and determine its eigenenergies $E$.
Taking the ansatz $\psi^{n}_j = C_{1} z_{1}^{j} + C_{2} z_{2}^{j}$ for the eigenstates, where $C_1$ and $C_2$ are constants, and performing the imaginary gauge transformation, we find that $z_{1}$ and $z_{2}$ are the roots of the eigenvalue equation $E = -J(z + z^{-1})$, with $z_{1} z_{2} = 1$ by Vieta's formula. 
To proceed, we set $z_{1} = e^{i\theta}$, where $\theta $ can be complex.
By imposing the boundary conditions $\psi_{0} = 0$ and $\psi_{L + 1} = \mu e^{\alpha L} \psi _1$, we obtain
\begin{equation}
    E = -2J \cos\theta, ~~\mathrm{where} ~~\sin\ab[\ab(L + 1)\theta] = \mu e^{\alpha L}\sin\theta.
    \label{Seq:E of H'}
\end{equation}
The eigenenergy becomes complex when $\mu e^{\alpha L} > 1$, from which we can estimate that the non-Hermitian skin effect (NHSE) is broken and the scale free localization (SFL) occurs at $\mu = e^{-\alpha L}$ when increasing $\mu $ from zero.

Next, by further increasing $\mu$, we determine the value of $\mu$ at which the approximation in Eq.~\eqref{Seq:H'} does not apply.
This corresponds to the condition, $\mu e^{-\alpha L} > 1$, from which we can estimate that the NHSE recovers at $\mu = e^{\alpha L}$.
The crossover can be understood as follows.
In this regime, the impurity bond strongly couples the two sites at the boundary.
Therefore, the model can be effectively regarded as a boundary dimer and an open chain with $L-2$ sites:
\begin{equation}
    H' \simeq -J \sum_{j = 2}^{L - 2} \ab(c_{j}^\dagger c_{j + 1} + c_{j + 1}^\dagger c_{j}) - \mu J \ab(e^{\alpha L} c_{L}^\dagger c_{1} + e^{-\alpha L} c_{1}^\dagger c_L).
\end{equation}
The chain under an effective OBC exhibits the NHSE, and its eigenenergies become real.
In addition, the impurity bond induces two isolated dimer modes with approximate eigenenergies $\pm \mu J$.

Moreover, to understand the behavior of the condition number in the NHSE and SFL regimes,
we analytically evaluate the localization length for representative values of $\mu$.
Since the NHSE occurs for $\mu \ll e^{-\alpha L}$ in the present model,
we consider the OBC limit, namely, $\mu =0$.
Then, we obtain $ \theta _n = {n \pi}/(L+1), ~(n=1,2, \cdots ,L)$ from Eq.~\eqref{Seq:E of H'}.
Because the original eigenstates of $H$ are given by $e^{-\alpha j}\psi ^n_j$ and $\theta _n$ is real,
the localization in the NHSE regime is given by $\xi = \alpha ^{-1}$.
The system also exhibits the NHSE for $\mu \gg e^{\alpha L}$.
In this case, the system can be seen as an open chain with a dimer.
Therefore, the chain shows the NHSE with the localization length $\xi =\alpha ^{-1}$ in the same way as in the OBC limit.
As a result, the condition number does not exhibit the dependence on $\mu$ in the NHSE regimes, as shown in Fig.~2 in the main text.

We also consider the regime where the SFL occurs for $\mu \ll 1$ using the approximated model in Eq.~\eqref{Seq:H'}.
The two boundary conditions require
\begin{equation} 
    (z_1^{L+1}-z_2^{L+1})=\mu e^{\alpha L}(z_1 - z_2).
    \label{eq:boundary condition in SFL}
\end{equation}
In this regime, we have complex energy eigenvalues satisfying $|z_{1}| > 1 > |z_{2}|$. 
Therefore, from $|z_1|^{L+1}\simeq \mu e^{\alpha L}|z_1|$  approximately derived from Eq.~\eqref{eq:boundary condition in SFL}, we obtain $|z_{1}| = e^{\alpha + \ln \mu / L}$. 
Consequently, the localization length of the original wavefunction is given by $\xi \simeq |\ln\mu|^{-1} L$.
As seen in Eq.~(26) in the Methods section, the condition number in this case is expected to be proportional to $e^{|\ln\mu|}$. This result is consistent with the numerical results obtained in Fig.~2{\bf b}.

\begin{figure}[t]
    \centering
    \includegraphics[width=\linewidth]{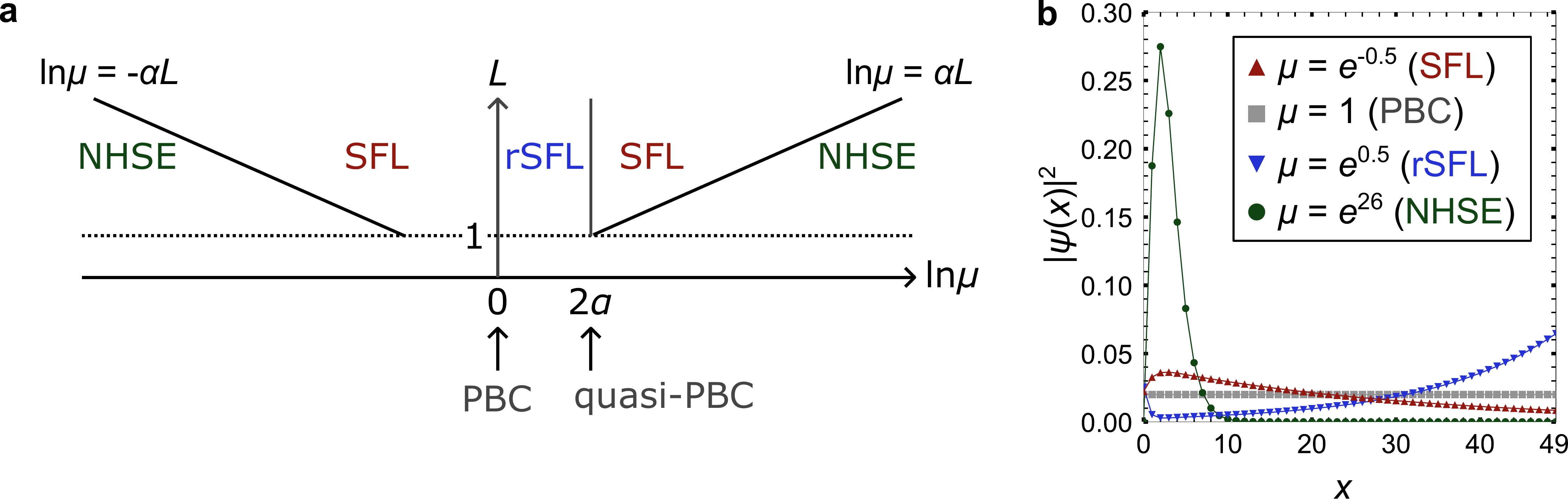}
    \caption{{\bf Regimes for the uniform system.} {\bf a} Regime diagram of the model~\eqref{Seq:imag. gauge transformed H} for $\lambda = 0$ in the $(\ln \mu, L)$ plane. {\bf b} Spatial distributions of the density in various regimes for  $\alpha = 0.5$ and $L = 50$.
    }
    \label{Sfig:simple model regime}
\end{figure}

The regime diagram of this uniform model has been obtained in Ref.~\cite{Li2021}. For illustration, here we present it in Supplementary Fig.~\ref{Sfig:simple model regime}{\bf a}, which includes the region for $\mu >1$.
The above estimation of the SFL-NHSE critical length is consistent with the previous numerical analysis.
In the intermediate region of $\mu$ ($e^{-2\alpha} < \mu < e^{2\alpha}$), the system exhibits an SFL regime localized near $j = 1$, similar to the NHSE regime, for $\mu < 1$, and a reversed SFL (rSFL) regime localized near the opposite edge at $j = L$ for $\mu > 1$.
At $\mu = e^{2\alpha}$, an extended regime, referred to as the quasi-periodic boundary conditions (quasi-PBC), emerges between the sharp transition
 between the sharp transition from the rSFL to the SFL regime.
Supplementary Figure~\ref{Sfig:simple model regime}{\bf b} shows the representative density profiles in each of the regimes.

In the main text, we consider the case $\lambda \neq 0$. When quasiperiodicity, non-reciprocity, and GBC coexist, the eigenstate with the smallest real part is typically dominated by an impurity-induced isolated mode. To gain insight into the localization properties, we therefore examine the wavefunction profile of the right eigenstate with the second smallest real part. In the complex spectrum, this state lies at the band edge and, in our model, always possesses a real eigenvalue.

\section{Supplementary Note 2: Characterizing regimes via the entanglement entropy}

In this section, we discuss the entanglement entropy (EE), 
which is defined for a given real-space many-body state and serves as a measure of how inhomogeneously that state is distributed.
When using EE as a quantity to characterize the system, it is appropriate to consider its value in the steady state reached after long-time evolution.

To proceed, we consider a system of $N$ free fermions, and denote the state at time $t$ by $\ket|\Psi_{t}>$.
The state $\ket|\Psi_{t}>$ is represented as a collection of $N$ mutually orthogonal single-particle state vectors.
For the initial state, we take a half-filled and evenly distributed state in the real-space representation, where particles occupy every other site:
\begin{equation}
    \ket|\Psi_{0}> = \prod_{j = 1}^{\floor{L / 2}} c_{2j - 1}^\dagger \ket|0> ,
    \label{Seq:Psi_0}
\end{equation}
with the vacuum state $\ket|0>$.
By applying the time-evolution operator $e^{iHt}$, we obtain the time-evolved state.

Since this is a fermionic system, the single-particle states within $\ket|\Psi_{t}>$ must remain orthogonal.
To preserve this orthogonality, we follow the procedure discussed in Ref.~\cite{Kawabata2023} to perform the time evolution.
Specifically, we express the state at time $t$ as
\begin{equation}
    \ket|\Psi_{t}> = \prod_{n = 1}^N \ab\{\sum_{j = 1}^L U_{jn}(t) c_{j}^\dagger\} \ket|0>,
    \label{Seq:Psi_t}
\end{equation}
where $U$ is an $L \times N$ matrix. For example, at $t = 0$, we have $U_{jn}(0) = \delta_{j, 2n - 1}$.
To evolve this state by a time step $\Delta t$, we compute $e^{iH\Delta t}U$, and then perform a QR decomposition,
\begin{equation}
    e^{iH\Delta t}U = QR,
    \label{Seq:QR decomp.}
\end{equation}
from which we obtain the new orthonormalized state $U(t + \Delta t) = Q$.

Since $U(t)$ represents the $N$-particle state at time $t$, the correlation matrix is given by
\begin{equation}
    C(t) = \ab[U(t) U^\dagger(t)]^T,
    \label{Seq:correlation matrix}
\end{equation}
which yields the correlation function $C_{jk}(t) = \braket<\Psi_{t} | c_j^\dagger c_k | \Psi_{t}>$. The superscript $T$ represents the transposition. 
We then restrict the correlation matrix to an $l \times l$ sub-block, corresponding to partitioning the system into subsystems of size $l$ and $L - l$.
By diagonalizing this block, we obtain the eigenvalues $\zeta_{j}(t)$, from which the EE at time $t$ is calculated as
\begin{equation}
    S_{l}(t) = -\sum_{j = 1}^l \ab[\zeta_{j}(t) \ln\zeta_{j}(t) + \ab\{1 - \zeta_{j}(t)\} \ln\ab\{1 - \zeta_{j}(t)\}].
    \label{Seq:EE}
\end{equation}
In our calculations, we set $\Delta t = 1$ and $l = L / 2$.
The steady-state value of $S_{L/2}$ was obtained by averaging its values over the interval $t = 1000$ to $2000$
to ensure that a  steady state is reached.

In systems with non-reciprocal hopping, the entanglement entropy $S_{L/2}$ follows a logarithmic law for extended or SFL regimes, an area law for localized or NHSE regimes, and a volume law for critical states, depending on the localization properties of the particles~\cite{Wang2023,Zhou2024,Li2024}.

We summarize our results in Supplementary Fig.~\ref{Sfig:EE}.
Taking several $\mu$ values as representatives, we observe that the crossover in the scaling behavior of the EE in each $\mu$ region coincides with the change in the scaling of the non-normality ratio presented in Fig.~2 in the main text and Supplementary Fig.~\ref{Sfig:cond. num with strong mu} (see below).
At the bulk localization transition point $\lambda = \lambda_{\mathrm{c}}$, the EE follows the volume law, while in the localized phase for $\lambda > \lambda_{\mathrm{c}}$, it obeys the area law.
For $\lambda < \lambda_{\mathrm{c}}$, Supplementary Figs.~\ref{Sfig:EE}{\bf a} show that EE follows the area law for small $L$ due to the NHSE and the logarithmic law for large $L$ due to the SFL, with the corresponding crossover length increasing with $\lambda$.
In Supplementary Figs.~\ref{Sfig:EE}{\bf b}, the EE follows a logarithmic law, consistent with the SFL~\cite{Wang2023} or the extended regime~\cite{Zhou2024,Li2024}.

\begin{figure}
    \centering
    \includegraphics[width=0.6\linewidth]{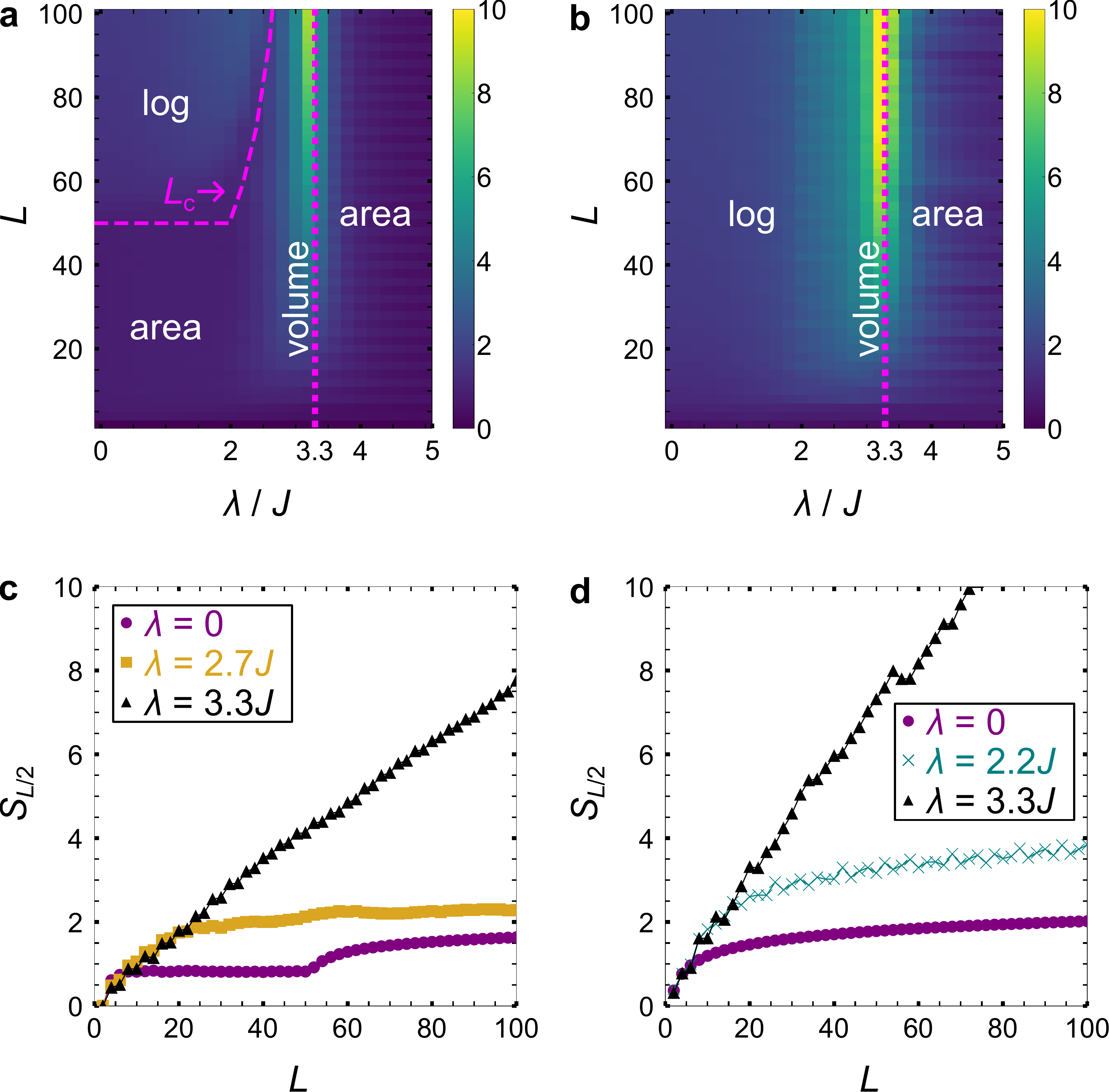}
    \caption{{\bf Regime diagrams based on entanglement entropy.} {\bf a},{\bf b} Entanglement-based regime diagrams under GBC. The color bar represents $S_{L/2}$. (a-2,b-2) Scaling behavior of the EE. The corresponding impurity strengths correspond to $\ln\mu = -25$ ({\bf a}) and $\ln\mu = -0.5$ ({\bf b}).
    }
    \label{Sfig:EE}
\end{figure}

\section{Supplementary Note 3: Regime diagram in the strong impurity regions}

In this section, we consider the case where the hopping across the impurity is stronger than other bonds and present Supplementary Fig.~\ref{Sfig:cond. num with strong mu}.
For $\ln\mu = 0.5$, which is close to the PBC limit, a crossover is found from the rSFL regime, where the non-normality ratio follows a logarithmic scaling law, to the extended regime, where it remains $O(1)$ as shown in Supplementary Fig.~\ref{Sfig:cond. num with strong mu}{\bf a}.
For large impurity strength ($\ln\mu = 25$), a regime diagram in Supplementary Fig.~\ref{Sfig:cond. num with strong mu}{\bf c} similar to Fig.~2{\bf c} in the main text is obtained, including the NHSE–SFL crossover.
In addition, as in Fig.~2{\bf c}, the critical length $L_{\mathrm{c}}$ (black dashed curve) agrees well with the value estimated from Fig.~2{\bf a}. 
However, the rSFL regime emerges for $\mu = 1$, as seen in Supplementary Fig.~\ref{Sfig:simple model regime}{\bf a}.
Therefore, the values of $\kappa_{\mathrm{R}}$ for $\ln\mu = -0.5$ and $\ln\mu = 0.5$ are not identical.

For the same values of $\mu$, Supplementary Figs.~\ref{Sfig:cond. num with strong mu}{\bf b},{\bf d} present the entanglement regime diagram and the scaling laws for several values of $\lambda$, in a manner similar to Supplementary Fig.~\ref{Sfig:EE}. As in the low-$\mu$ case shown in Supplementary Fig.~\ref{Sfig:EE}, the regime diagrams are consistent with those derived from the condition number in Supplementary Figs.~\ref{Sfig:cond. num with strong mu}{\bf a} and \ref{Sfig:cond. num with strong mu}{\bf c}, respectively.

In summary, the non-normality ratio introduced here can serve as a measure to construct regime diagrams, complementing the EE, as demonstrated in Fig.~2 in the main text and Supplementary Fig.~\ref{Sfig:cond. num with strong mu}.
We additionally remark that the non-normality ratio has a practical advantage, as it is much less numerically demanding than the EE while it can reliably distinguish among the SFL, NHSE, and extended regimes.

\begin{figure}[t]
    \centering
    \includegraphics[width=\linewidth]{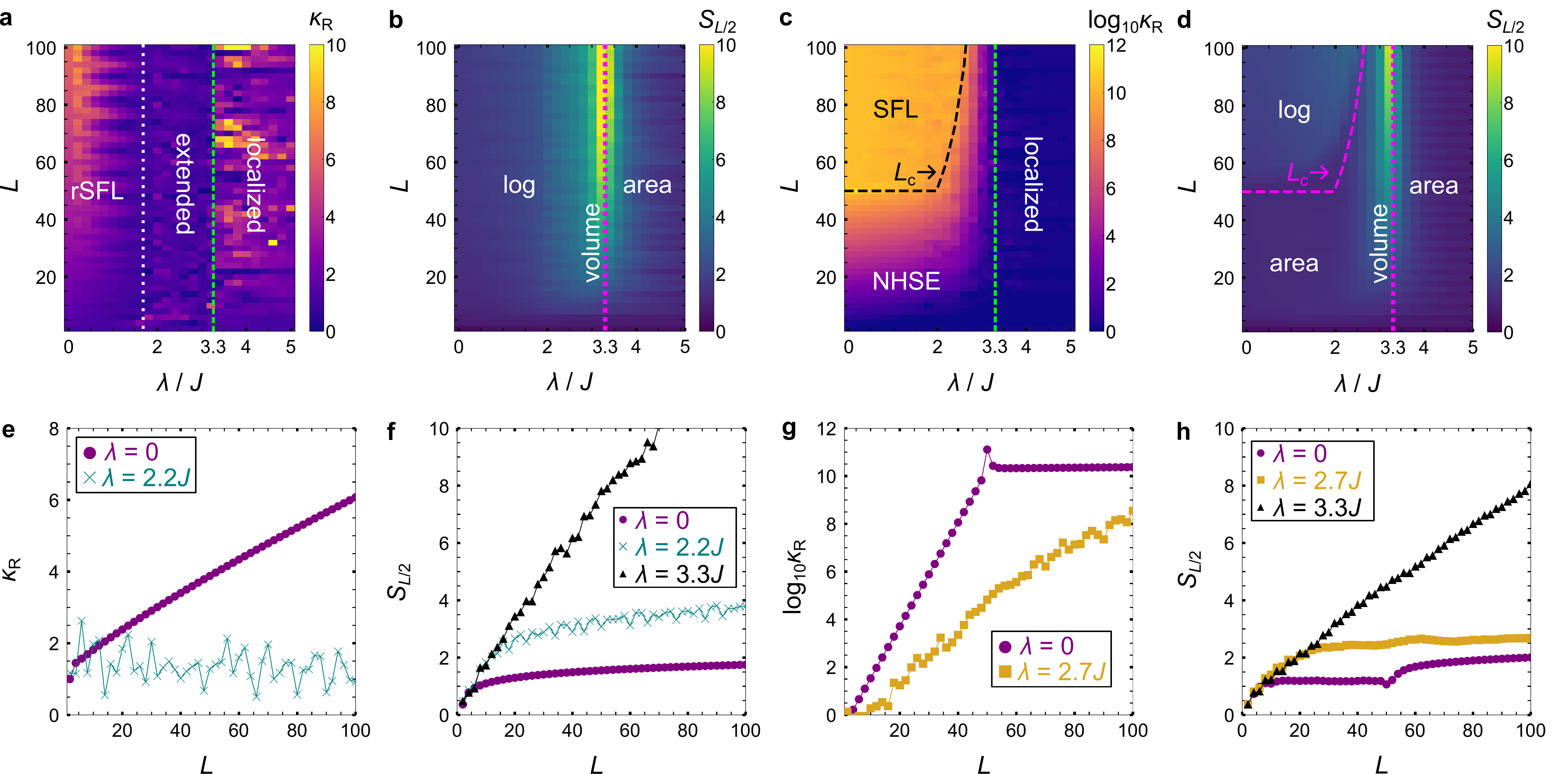}
    \caption{{\bf Regime diagrams in regions for $\mu > 1$.} {\bf a},{\bf c} Non-normality ratio-based regime diagrams, {\bf b},{\bf d} EE-based regime diagrams on the $(\lambda, L)$ plane and {\bf e},{\bf g} Scaling behavior of the non-normality ratio $\kappa_{\mathrm{R}}$ and {\bf f},{\bf h} of the EE $S_{L/2}$ for $\ln \mu = 0.5$ ({\bf a},{\bf b},{\bf e},{\bf f}) and $\ln \mu = 25$ ({\bf c},{\bf d},{\bf g},{\bf h}). The dashed curve in {\bf c},{\bf d} marks the critical length $L_{\mathrm{c}}$ estimated from Fig.~2{\bf a} as in Fig.~2{\bf b}~and~Fig.~4{\bf a}.}
    \label{Sfig:cond. num with strong mu}
\end{figure}

\section{Supplementary Note 4: Disconnection of an impurity bond via quasiperiodicity}

In this section, we analyze how quasiperiodicity suppresses hoppings across the impurity bond and in the bulk.
To verify that the quasiperiodicity decouples the impurity bond links prior to the bulk links, we examine the ratio of their expectation values defined as:
\begin{equation}
    \frac{ \left| \mu J\braket<e^{\alpha} c_{1}^\dagger c_{L} + e^{-\alpha} c_{L}^\dagger c_{1}> \right|}
    {\left| J\braket<e^{\alpha} c_{\text{bulk}}^\dagger c_{\text{bulk} + 1} + e^{-\alpha} c_{\text{bulk} + 1}^\dagger c_{\text{bulk}}> \right|}.
    \label{Seq:ratio btw. boundary and bulk hop}
\end{equation}
Here, we define the expectation value of the operator $\hat{O}$ as
\begin{equation}
    \braket<\hat{O}> \coloneq \frac{1}{L} \sum_{n = 1}^{L} \lbra{E_n}\hat{O}\ket|E_n>,
    \label{Seq:def. of expectation val.}
\end{equation}
where $\ket|E_n>$ ($\lket{E_n}$) denotes the $n$-th right (left) eigenstate. 
The calculated results are shown in Supplementary Fig.~\ref{Sfig:disconnection}. The behavior of this ratio confirms that the hopping across the impurity bond is effectively decoupled for $\lambda < \lambda_{\mathrm{c}}$. The corresponding value of $\lambda$ is consistent with the estimate ($\lambda \approx 2.6J$) obtained from Fig.~2{\bf a} for $\alpha = 0.5$, $L = 89$ and $\ln\mu = -25$. 
For this parameter set, we have $\lambda_{\mathrm{c}} \approx 3.3J$. 
This conclusion remains robust regardless of the choice of other bonds, provided they are located sufficiently deep within the bulk.

\begin{figure}
    \centering
    \includegraphics[width=0.5\linewidth]{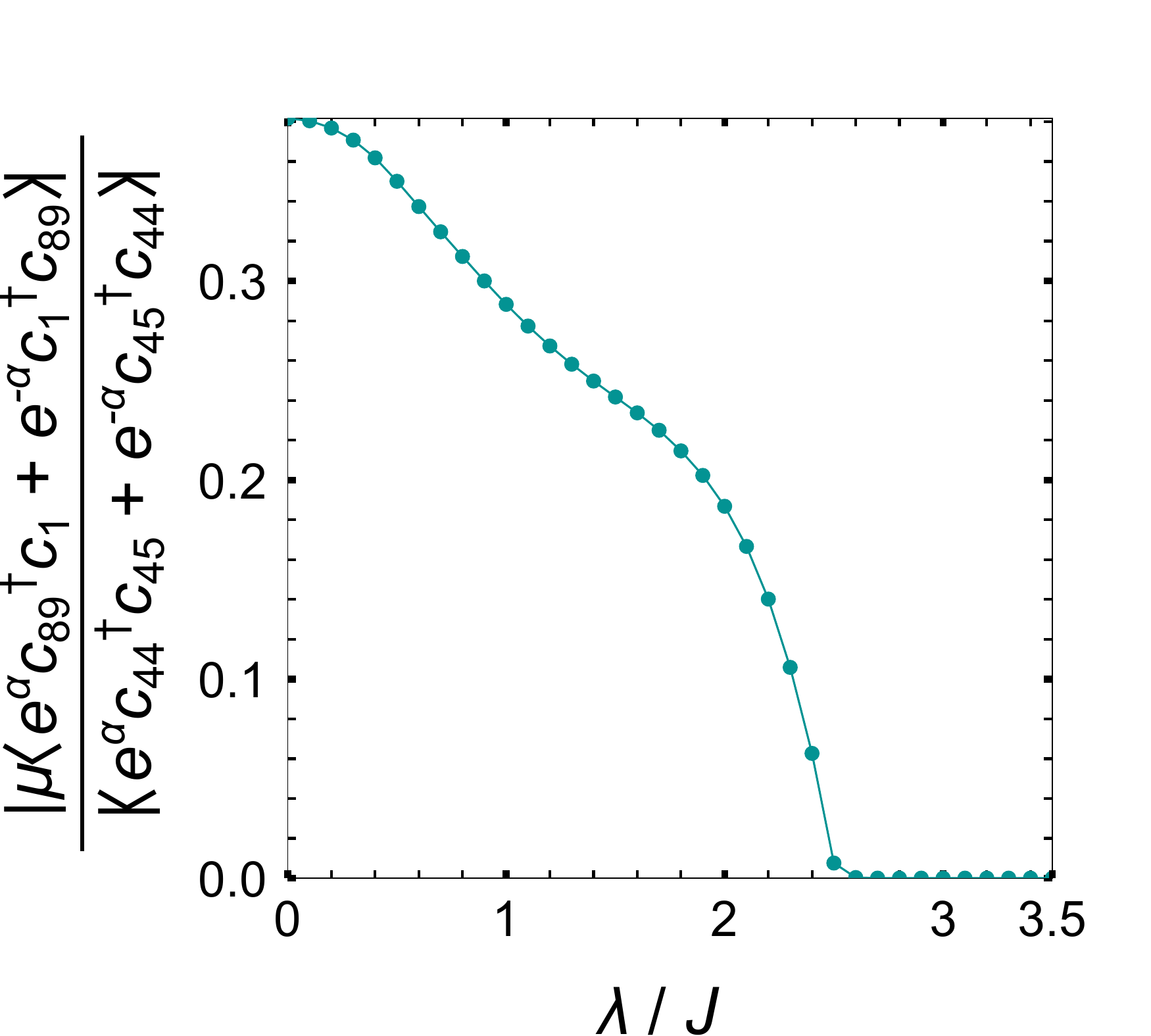}
    \caption{{\bf Impurity-bond amplitude.} Ratio of the absolute value of the boundary hopping expectation to that of the bulk hopping as a function of the quasiperiodic potential strength $\lambda$.
    The bulk hopping is evaluated between the $\floor{L / 2}$-th and $(\floor{L / 2} + 1)$-th sites.
    Here we adopt the same parameter set as in Fig.~3 in the main text, leading to $\lambda_{\mathrm{c}} \approx 3.3J$. 
    }
    \label{Sfig:disconnection}
\end{figure}

\end{bibunit}
\end{widetext}

\end{document}